\documentstyle[aps,prl,multicol]{revtex}

\begin{document}
\draft
\title{Localization transition of random copolymers at interfaces}
\author{Semjon Stepanow $^a$, Jens-Uwe Sommer$^b$, and Igor Ya. Erukhimovich$^c$}
\address{$^a$ Universit\"at Halle, Fachbereich Physik, D-06099 Halle/Saale, Germany}
\address{$^b$ Universit\"at Freiburg, Theoretische Polymerphysik, Rheinstra\ss e\\
12,D-79104 Freiburg, Germany}
\address{$^c$ Moscow State University, Physics Department, Moscow 117234, Russia}
\date{\today }
\maketitle

\begin{abstract}
We consider adsorption of random copolymer chains onto an interface within
the model of Garel et al. Europhysics Letters {\bf 8,} 9 (1989). By using
the replica method the adsorption of the copolymer at the interface is
mapped onto the problem of finding the ground state of a quantum mechanical
Hamiltonian. To study this ground state we introduce a novel variational
principle for the Green's function, which generalizes the well-known
Rayleigh-Ritz method of Quantum Mechanics to nonstationary states.
Minimization with an appropriate trial Green's function enables us to find
the phase diagram for the localization-delocalization transition for an
ideal random copolymer at the interface. 
\end{abstract}


\widetext
The presence of copolymers at interfaces between two immiscible fluids is
crucial to such processes as emulsion stabilization, wetting, microemulsion
formation and reinforcement of polymer-polymer interfaces \cite{hancock}-
\cite{milner}. For oil-water interfaces, the polymers that are used in these
applications are amphiphilic in nature. While one component is soluble in
the oil phase, the other component is soluble in water. The difference in
solubilities drives the copolymers to adsorb at the interface between the
two phases. The localized copolymers can stabilize the interface in the
sense that they significantly reduce the surface tension. The study of
random copolymers has been motivated in recent years by relevance of these
materials for both biological and technological applications. Moreover, the
properties of these simple random systems may be important in understanding
of much more complex systems such as proteins \cite{karplus}-\cite{sali}.

It is well-known that adsorption of long polymer chains on surfaces or
interfaces is related to the bound state problem of a Quantum Mechanical
(QM) particle in a potential well \cite{degennes}. The Green's function $G(%
{\bf r},{\bf r}^{\prime };s,s^{\prime })$ of a chain in $d$ spatial
dimensions with the monomer $s$ at position ${\bf r}$ and the monomer $%
s^{\prime }$ at ${\bf r}^{\prime }$ is a fundamental quantity in the
statistical mechanics of polymers. It obeys in the presence of an external
potential $V({\bf r},s)$ the following equation

\begin{equation}
\frac{\partial }{\partial s}G=\frac{l^{2}}{2d}\nabla ^{2}G-\frac{V}{kT}G,
\label{greensfunktion}
\end{equation}
with the condition: $G({\bf r},{\bf r}^{\prime },0)=\delta ({\bf r}-{\bf r}%
^{\prime })$ and $l$ being the statistical segment length of the polymer.
Eq.(\ref{greensfunktion}) is related to the Schr\"{o}dinger equation by
using the replacement: $s\rightarrow it$, $l^{2}/(dkT)\rightarrow 1/m$, $%
kT\rightarrow \hbar $ \cite{edwards},\cite{degennes}. Considering an
external potential independent of the monomer species along the chain and
given by $V(x)/kT=-u\delta (x)$ in one spatial dimension, a localized part
of the solution exist and can be written as $G_{loc}(x,x^{\prime },N)=k\exp
(l^{2}Nk^{2}/2))\exp (-k|x|-k|x^{\prime }|)$. It becomes the only relevant
contribution for $N=|s-s^{\prime }|\rightarrow \infty $. Here the
localization length $\xi =1/k$ is given by $\xi =1/u$.

However, random copolymers utilized in order to reinforce polymer-polymer
interfaces do not adsorb according to the above scenario. The external
potential in that case has an arc-length dependence, which in QM picture
corresponds to a time dependent potential, so that an appearance of a bound
state is not so obvious. In contrast to the example considered above, the
individual monomers belonging to a random copolymer are {\it not a priori}
attracted by the interface. The localization effect appears to be due to the
feed-back of the interaction of larger parts of the chain with the
interface. Garel et al. \cite{garel} introduced a simple model describing
the behavior of an ideal copolymer chain in the presence of an interface
between two phases by the following Edwards functional integral

\begin{equation}
Z=\int {\cal D}[x(s)]\exp \{-\frac{1}{2l^{2}}\int_{0}^{N}ds(\partial
x/\partial s)^{2}+w\int_{0}^{N}ds\zeta (s){\rm sgn}(x(s))\},  \label{ad1}
\end{equation}
where $\zeta (s)$ is a random Gaussian variable describing the heterogeneity
of the copolymer. Due to the fact that there are no interactions between the
monomers, the coordinates $y(s)$ and $z(s)$ along the interface separate, so
that the problem becomes one-dimensional. In the model of Garel et al. \cite
{garel} the distribution function of the random variable $\zeta (s)$ reads
for the discrete version of the polymer chain ($x(s)\rightarrow x_{i}$, $%
\zeta (s)\rightarrow \zeta _{i}$, $i=1,...,N$ with $N$ being the degree of
polymerization of the chain) $P(\zeta _{i})=(2\pi \Delta _{0})^{-1/2}\exp
(-(\zeta _{i}-\zeta _{0})^{2}/2\Delta _{0})$, where $\zeta _{0}$ describes
the asymmetry in the composition of the copolymer. The average over $\zeta
_{i}$ within the replica method \cite{garel} leads to the following
Replica-Hamiltonian

\begin{equation}
H_{n}=-D\sum_{\alpha =1}^{n}\nabla _{x}^{2}-w\zeta _{0}\sum_{\alpha =1}^{n}%
{\rm sgn}(x_{\alpha })-\frac{1}{2}\Delta _{0}w^{2}\sum_{\alpha ,\beta =1}^{n}%
{\rm sgn}(x_{\alpha }){\rm sgn}(x_{\beta }),  \label{ad3}
\end{equation}
where $n$ is the number of replicas and $D$ is given by $D=l^{2}/2$. So far,
besides Imry-Ma type arguments \cite{garel,spb} there is no analytical
treatment enabling one to find the ground state of Hamiltonian $H_{n}$,
which is related to the localization of the random copolymer. The Hartree
method applied in Ref.\cite{garel} fails to describe the localized state.
The problem at hand it also of general interest since it is a further
example where frozen-in disorder (here chemical disorder) induces
localization of the chain.

In this Letter we present a novel variational principle in order to study
the ground state of the Hamiltonian $H_{n}$ given in Eq.(\ref{ad3}). In
contrast to the well-known stationary Rayleigh-Ritz (RR) method \cite{landau}%
, where the function of interest is the wave function of the problem, we are
interested in a variational principle for the full Green's function (GF).
While the GF is a dynamic quantity (in the QM picture), the latter is a
generalization of RR extremal principle to nonstationary states. We note
that our variational principle for the Green's function is not restricted to
treat the Hamiltonian $H_{n}$, Eq.(\ref{ad3}), but it is in fact a general
method of quantum mechanics and in particular is a promising method for
treating polymer problems.

In order to introduce the variational principle we start with the definition
of the GF written in the form 
\begin{equation}
-G^{-1}(z)+G_{0}^{-1}(z)+H_{i}=0,  \label{ad4}
\end{equation}
where $G_{0}$ is the Greens function without interaction, $%
G_{0}^{-1}(z)=z+H_{0}$, $H_{0}$ is the unperturbed part of the Hamiltonian, $%
z$ is Laplace conjugate to the chain's length (time in QM), and $H_{i}$ is
the interaction part of the Hamiltonian. Considering Eq.(\ref{ad4}) as a
stationarity condition, $\delta {\cal F}(G)/\delta G=0$, of a functional $%
{\cal F}(G)$, we obtain 
\begin{equation}
{\cal F}(G)=-{\rm tr}\ln (G)+{\rm tr}G_{0}^{-1}G+{\rm tr}H_{i}G.  \label{ad5}
\end{equation}
Eq.(\ref{ad5})\ defines an extremum functional, the stationarity condition
of which yields the exact Green's function. We now present a refined version
of this principle. Iterating Eq.(\ref{ad4}) we arrive at 
\begin{equation}
-(1-H_{i}G_{0})G^{-1}+G_{0}^{-1}-H_{i}G_{0}H_{i}=0.  \label{ad6}
\end{equation}
Considering the latter as a stationarity condition for a functional ${\cal F}%
^{\prime }(G)$, we get 
\begin{equation}
{\cal F}^{\prime }(G)=-{\rm tr}(1-H_{i}G_{0})\ln (G)+{\rm tr}G_{0}^{-1}G-%
{\rm tr}H_{i}G_{0}H_{i}G.  \label{ad7}
\end{equation}
For practical purpose it may be useful to start with ${\cal F}^{\prime }$
instead of ${\cal F}$. We remark that ${\cal F}(G)$ and especially ${\cal F}%
^{\prime }(G)$ are similar to the generating functional of the 2nd Legendre
transform in statistical physics and in field theory \cite{dedominicis}-\cite
{erukhimovich}. It is amazing that the method based on (\ref{ad5}, \ref{ad7}%
) has been discovered only now.

We now will apply the above method to find the ground state of an asymmetric
localization potential given by $U(x)=-u\delta (x)+\chi \theta (x),$ where $%
\theta (x)$ is the step function. Assuming ground state dominance (GSD) \cite
{degennes} we choose the trial Green's function as

\begin{eqnarray}
G(k_{1},k_{2};x,x^{\prime },t) &=&\frac{2k_{1}k_{2}}{k_{1}+k_{2}}\exp
(-\varepsilon _{k}t)(\exp (-k_{1}x)\theta (x)+\exp (k_{2}x)\theta (-x)) 
\nonumber \\
&&\ (\exp (-k_{1}x^{\prime })\theta (x^{\prime })+\exp (k_{2}x^{\prime
})\theta (-x^{\prime })),  \label{ad9}
\end{eqnarray}
where $\varepsilon _{k}=-Dk_{2}^{2}$, and $k_{1}$ and $k_{2}$ are fit
parameters. Inserting $G$ into ${\cal F}(G)$ yields

\begin{equation}
{\cal F}(k_{1},k_{2})=\ln (z+\varepsilon _{k})+\frac{-\varepsilon
_{k}+Dk_{1}k_{2}}{z+\varepsilon _{k}}-\ \frac{u}{z+\varepsilon _{k}}\frac{%
2k_{1}k_{2}}{k_{1}+k_{2}}+\frac{\chi }{z+\varepsilon _{k}}\frac{k_{2}}{%
k_{1}+k_{2}}.  \label{ad10}
\end{equation}

Solving equations associated with stationarity conditions 
\begin{equation}
\partial {\cal F}/\partial k_{1}=\partial {\cal F}/\partial k_{2}=0
\label{ad11}
\end{equation}
we get the fit parameters $k_{1}$ and $k_{2}$ as 
\begin{equation}
k_{1}=\frac{1}{2}(\frac{u}{D}+\frac{\chi }{u}),\,\,\,\,\,\,\,k_{2}=\frac{1}{2%
}(\frac{u}{D}-\frac{\chi }{u}).\,\,\,\,  \label{ad12}
\end{equation}
The latter coincide with those obtained from the exact solution of the
Schr\"{o}dinger equation in the asymmetric potential. For disappearing
asymmetry ($\chi =0)$ the ground state will become symmetric ($k_{1}=k_{2}$%
). The delocalization transition is defined by the condition $k_{2}=0$. We
note that the minimization procedure based on (\ref{ad5}) gives the same
result as the Rayleigh-Ritz approach. We can also consider $\varepsilon _{k}$
in Eq.(\ref{ad10}) as a free parameter and minimize ${\cal F}(\varepsilon
,k_{1},k_{2})$ given by Eq.(\ref{ad10}) with respect to $\varepsilon $, $%
k_{1}$ and $k_{2}$. The stationarity condition with respect to $\varepsilon $
fixes $\varepsilon $ as the expectation value of the Hamiltonian associated
with (\ref{ad10}). The stationarity conditions with respect to $k_{1}$ and $%
k_{2}$ coincide exactly with those obtained by using the Rayleigh-Ritz
method.

We now turn to the adsorption of a random copolymer chain, or what is
equivalent to the study of the ground state of the Hamiltonian ${\cal H}_{n}$%
, Eq.(\ref{ad3}). In the framework of the variational principle based on the
functional ${\cal F}$, which is given by Eq.(\ref{ad5}), the simplest $n$%
-replica trial Green's function is

\begin{eqnarray}
G(k_{1},k_{2};x,x^{\prime },t) &=&(\frac{2k_{1}k_{2}}{k_{1}+k_{2}})^{n}\exp
(-n\varepsilon _{k}t)\prod\limits_{\alpha =1}^{n}((\exp (-k_{1}x_{\alpha
})\theta (x_{\alpha })+\exp (k_{2}x_{\alpha })\theta (-x_{\alpha })) 
\nonumber \\
&&(\exp (-k_{1}x_{\alpha }^{\prime })\theta (x_{\alpha }^{\prime })+\exp
(k_{2}x_{\alpha }^{\prime })\theta (-x_{\alpha }^{\prime })))  \label{ad13}
\end{eqnarray}
with $\varepsilon _{k}=-Dk_{2}^{2}$ in the case $\zeta _{0}>0$. The energy $%
\varepsilon _{k}$ is negative due to the same reason as in the case of the
localization of a QM particle. Notice that the trial function in Eq.(\ref
{ad13}) is a product over the one replica Green's functions. This is due to
the circumstance that the main effect of the randomness here is expected to
result in an attraction of the copolymer to the interface. In the case when
the randomness would result in self-interactions between the monomers, a
product over replica pairs would be more appropriate. Eq.(\ref{ad13}) can be
considered as the first term in the expansion of the $n$-replica Green's
function over one-replica, two-replica, ... Green's functions. The
evaluation of ${\cal F}$ gives

\begin{equation}
{\cal F}(k_{1},k_{2})=\ln (z+n\varepsilon _{k})+\frac{-n\varepsilon
_{k}+nDk_{1}k_{2}}{z+n\varepsilon _{k}}-\frac{w\zeta _{0}n}{z+n\varepsilon
_{k}}\frac{k_{2}-k_{1}}{k_{1}+k_{2}}-\frac{n}{2}\Delta _{0}w^{2}\frac{1}{%
z+n\varepsilon _{k}}-\frac{n(n-1)}{2}\Delta _{0}w^{2}\frac{1}{z+n\varepsilon
_{k}}(\frac{k_{2}-k_{1}}{k_{1}+k_{2}})^{2}.  \label{ad14}
\end{equation}

After some algebra the stationarity conditions (\ref{ad11}) associated with (%
\ref{ad14}) simplify in the limit $n=0$ to 
\begin{equation}
k_{2}^{3}+3\,k_{2}^{2}k_{1}+3\,k_{2}\,k_{1}^{2}+k_{1}^{3}+2\,\zeta
\,k_{2}+2\,\zeta \,k_{1}-2\,\Delta \,k_{2}+2\,\Delta \,k_{1}=0,  \label{ad15}
\end{equation}

\begin{equation}
k_{2}^{4}+3\,k_{2}^{3}k_{1}-k_{2}^{2}\zeta +3\,k_{2}^{2}k_{1}^{2}-2\,\Delta
\,k_{1}\,k_{2}+k_{2}\,k_{1}^{3}+\zeta \,k_{1}^{2}=0,  \label{ad16}
\end{equation}
where the quantities $\Delta =\Delta _{0}w^{2}/D$ and $\zeta =\zeta _{0}w/D$
are introduced. The solution for $k_{1}$ and $k_{2}$ is obtained as $%
k_{2}=ak_{1}$, where $a$ is given by $a=(\zeta +2\,\Delta +S)/(2\,\Delta
-3\,\zeta )$ with $S=\sqrt{4\,\zeta ^{2}+2\,\Delta \,\zeta +4\,\Delta ^{2}{}}
$. The quantities $k_{1}$ and $k_{2}$ are obtained from Eqs.(\ref{ad15}-\ref
{ad16}) for $\zeta >0$ as 
\begin{equation}
k_{1}=\sqrt{2}\left( 2\,\Delta -3\,\zeta \right) {\frac{(2\,\zeta ^{2}-\zeta
\,S+\Delta \,S)^{1/2}}{\left( -2\,\zeta +4\,\Delta +S\right) ^{3/2}}}
\label{ad18}
\end{equation}
\begin{equation}
k_{2}=\sqrt{2}\left( \zeta +2\,\Delta +S\right) {\frac{(2\,\zeta ^{2}-\zeta
\,S+\Delta \,S)^{1/2}}{\left( -2\,\zeta +4\,\Delta +S\right) ^{3/2}}}
\label{ad19}
\end{equation}
The localization-delocalization transition occurs at the condition $%
2\,\Delta -3\,\zeta =0$, where $k_{1}=0$, i.e. the polymer delocalizes in
the right half-plane. The localization length $\xi =1/k_{1}$ becomes
infinite at the transition. Notice that $k_{2}$ (related to the chain's
penetration length into the unfavorable solvent) stays finite at the
transition. Using the relation for the energy per replica, $\varepsilon
_{k}=-Dk_{2}^{2}$, with $k_{2}$ given by Eq.(\ref{ad19}) we see that the
energy at the localization-delocalization transition is finite and is equal
to $-2/3\Delta $. This is due to the circumstance that in contrast to the
localization of a QM particle in a potential well, where the energy at the
transition is zero, the interaction of the stochastic copolymer with the
interface is in fact nonlocal.. The solution for $\zeta <0$ can be obtained
by using Eq.(\ref{ad14}) with $\varepsilon _{k}=-Dk_{1}^{2}$.

Considering now the symmetric situation ($\zeta _{0}=0$). From (\ref{ad18}-%
\ref{ad19}) we get 
\begin{equation}
k_{1}=\frac{\sqrt{6}}{9}\sqrt{\Delta },\,\,\,\,k_{2}=\frac{2\sqrt{6}}{9}\,%
\sqrt{\Delta }\,\,.  \label{ad20}
\end{equation}
Eq.(\ref{ad20}) tells us that the ground state breaks down the symmetry of
the Hamiltonian ${\cal H}_{n}$ with respect to the reflection $%
x\leftrightarrow -x$. An explanation can be given as follows: The condition $%
\zeta _{0}=0$ means that the fraction of the different types of monomers is
on average balanced. However, the symmetry of the composition is true only
for an ensemble of copolymers in the space of the quenched variable $\zeta
(s)$. The typical copolymer in such an ensemble contains an excess of
species of one type, which causes the asymmetry of the localized state. It
is remarkable that the Hamiltonian ${\cal H}_{n}$ contains information on
this asymmetry and that our variational approach is able to reflect this as
an asymmetry of the ground state.

Let us now compare the extremal principle based on the functional ${\cal F}%
(G)$ given by Eq.(\ref{ad5}) with that of Rayleigh-Ritz. The latter is based
on the minimization of the expectation value of the Hamiltonian $H_{n}$, $%
<k\mid $ $H_{n}\mid k>$, computed by using the trial function associated
with (\ref{ad13}). As in the case of localization of a particle in an
asymmetric potential well considered above, the minimization of (\ref{ad14})
with respect to $\varepsilon $, $k_{1}$ and $k_{2}$ completely coincides in
the limit of large $t$ with the Rayleigh-Ritz procedure. The difference
between both methods appears to be due to the condition $\varepsilon
_{k}=-Dk_{2}^{2}$. A restriction in the search of the bound state, which is
imposed by this condition, results apparently in selecting the states
breaking the reflection symmetry of the Hamiltonian $H_{n}$. This explains
why in contrast to the Rayleigh-Ritz method our approach is successful in
finding the localized state of the problem under consideration. While Eq.(%
\ref{ad15}) still coincides with the stationarity condition of $<k\mid
H_{n}\mid k>$ with respect to $k_{1}$, the left-hand side of Eq.(\ref{ad16})
differs from the derivative of $<k\mid H_{n}\mid k>$ with respect to $k_{2}$%
. The inspection of the condition $\partial {\cal F}/\partial k_{2}=0$ shows
that the latter provides the equality $\varepsilon _{k}=\,<k\mid H_{n}\mid
k>/n$. This shows that the minimization machinery in both variational
principle based on ${\cal F}(G)$ and that of Rayleigh-Ritz is in general
different, so that even in the stationary case the both extremal principles
are not completely equivalent with each other. In the example of the
localization of a QM particle in a potential well the both methods give,
however, the same result. The equivalence of Eq.(\ref{ad15}) with $\partial
<k\mid H_{n}\mid k>\partial k_{1}=0$ guarantees that the energy $\varepsilon
_{k}$ is extremal with respect to $k_{1}$.

We have found that the energy associated with the asymmetric solution (\ref
{ad20}) is lower than the energy obtained from the minimization of Eq.(\ref
{ad14}) in the symmetric case $k_{1}=k_{2}$. The computation of the inverse
Laplace transform of (\ref{ad14}) with respect to $z$ gives the functional $%
{\cal F}$ at the extremum as: ${\cal F}(t)=-\exp (Dnk_{2}^{2}t)/t$. Thus,
the asymmetric solution (\ref{ad20}) gives for all nonzero $n$ lower values
for$\ {\cal F}(t)$ than the symmetric one.

To conclude, we have studied the adsorption of an ideal random copolymer
chain at a selective interface. In terms of the replica method adsorption is
related to the ground state of a replica Hamiltonian. To proceed we have
introduced a novel variational principle for the Green's function. By using
an appropriate trial Green's function as a possible candidate for the ground
state we find the phase diagram of the localization-delocalization
transition for a random copolymer chain on the interface. We predict that
even in the case of a symmetric composition ($\zeta _{0}=0$) the ground
state is obtained to be asymmetric. This means that the ground state breaks
down the reflection symmetry of the Hamiltonian.

\acknowledgments Support from the Deutsche Forschungsgemeinschaft (SFB 418)
and from the Fond der Chemischen Industrie (FCI) as well as stimulating
discussions with H. Orland and A. Blumen are gratefully acknowledged.


\begin{references}
\bibitem{hancock}  R. I. Hancock, Surfactants; Th. F. Tadros, Ed. Academic
Press, New York, 1984.

\bibitem{brown}  H. R. Brown, V. R. Deline, and P. F. Green, Nature (London) 
{\bf 341}, 221 (1989).

\bibitem{brown1}  H. R. Brown, K. Char, V. R. Deline, and P. F. Green,
Macromolecules {\bf 26}, 4155 (1993); K. Char, H. R. Brown, , V. R. Deline,
Macromolecules {\bf 26}, 4164 (1993).

\bibitem{balazs}  C. Yeung, A. Balazs, D. Jasnow, Macromolecules {\bf 25},
1357 (1992).

\bibitem{spb}  J.-U. Sommer, G. Peng and A. Blumen, J. Chem. Phys. {\bf 105}%
, 8376 (1996).

\bibitem{dai}  C.-A. Dai, B. J. Dair, K. H. Dai, C. K. Ober, E. J. Kramer,
C.-Y. Hui, L. W. Jelinski, Phys. Rev. Lett. {\bf 73}, 2472 (1994).

\bibitem{noolandi}  J. Noolandi and A.-C. Shi, Phys. Rev. Lett. {\bf 74},
2836 (1995).

\bibitem{dai-1}  C.-A. Dai, B. J. Dair, K. H. Dai, C. K. Ober, E. J. Kramer,
C.-Y. Hui, L. W. Jelinski, Phys. Rev. Lett. {\bf 74}, 2837 (1995).

\bibitem{sommer}  J.-U. Sommer and M. Daoud, Europhysics Letters {\bf 32},
407 (1995).

\bibitem{milner}  S. Milner and G. H. Fredrickson, Macromolecules {\bf 28},
7953 (1995).

\bibitem{karplus}  M. Karplus and E. I. Shakhnovich, {\it in Protein Folding}%
, ed. by T. E. Creighton (Freeman, New York, 1995).

\bibitem{bryngelson}  J. Bryngelson and P. G. Wolynes, Proc. Natl. Acad.
Sci. U.S.A. {\bf 84}, 7524 (1987).

\bibitem{sali}  A. Sali et al., Nature (London) {\bf 369}, 248 (1994).

\bibitem{degennes}  P.G. de Gennes, Rep. Prog. Phys. {\bf 32}, 187, 1969.

\bibitem{edwards}  S. F. Edwards, Proc. Phys. Soc. London {\bf 85}, 613
(1965).

\bibitem{garel}  T. Garel, D. A. Huse, S. Leibler, H. Orland, Europhysics
Letters {\bf 8}, 9 (1989) .

\bibitem{dedominicis}  C. de Dominicis and P. C. Martin, J. Math. Phys. {\bf %
5}, 14 (1964); {\bf 5}, 31 (1964).

\bibitem{cornwall}  J. M. Cornwall, R. Jackiw, and E. Tomboulis, Physical
Review D {\bf 10}, 2428 (1974).

\bibitem{erukhimovich}  A. V. Dobrynin and I. Ya. Erukhimovich, J. Phys. II
France {\bf 1}, 1387 (1991).

\bibitem{landau}  L. D. Landau and E. M. Lifschitz, {\it Quantum Mechanics}
(Nauka, Moscow, 1974).
\end{references}
\end{document}